\documentstyle[preprint,revtex]{aps}
\begin{document}
\draft
\preprint{October, 1995}
\begin{title}
Exactly solvable extended Hubbard model
\end{title}
\author{D. F. Wang}
\begin{instit}
Institut de Physique Th\'eorique\\
Ecole Polytechnique F\'ed\'erale de Lausanne\\
PHB-Ecublens, CH-1015 Lausanne, Switzerland
\end{instit}
\begin{abstract}
In this work, we introduce long range version of the extended Hubbard
model. The system is defined on a non-uniform lattice.
We show that the system is integrable. The ground state,
the ground state energy, and the energy spectrum
are also found for the system. Another long range
version of the extended Hubbard model is also introduced
on a uniform lattice, and this system is proven to be integrable.
\end{abstract}
\narrowtext

In recent years, there have been considerable interests in
low dimensional electronic systems of low dimensions.
Systems of this type may exhibit interesting novel physics,
due to low dimensionality and strong correlation.
Anderson has suggested that the 2D one band Hubbard model
should explain the basic physics of the high temperature
superconductivity\cite{anderson}. It is suggested that the normal
state of the new cuperate oxide superconductors may
share the feature (Luttinger-liquid-like)
of 1D interacting electron gas\cite{anderson,su}.
This physical motivation
has initiated lots of recent activities in
one dimensional electronic models.

In one dimension, the Hubbard model was exactly solved
by Lieb and Wu\cite{wu}. With the $SO(4)$ symmetry due to Yang and
Zhang\cite{yang1,yang2}, the completeness
of the eigenstates in this structure has been discussed\cite{korepin1}.
More recently, an extended Hubbard model was introduced by
several authors in general $d$ dimensions\cite{korepin1}. By using
$\eta$-paring mechanism, they construct eigenstates
of the system, with off-diagonal-long-range order\cite{korepin2}.
Particularly, its one dimensional version was solved
exactly with Bethe-ansatz\cite{korepin2}.
In this work, we introduce
the long range version of the extended Hubbard model in
one dimension. We show that the system is completely integrable,
with explicit construction of infinite constants of motion.
The ground state, the energy spectrum
are found for the system.

Let us consider a one dimension lattice of sites $L$.
The positions of the sites are given by the roots of the
Hermite polynomial $H_L(x)$.
The roots of the Hermite polynomial
$r_1, r_2, \cdots, r_L$ are all real and distinct,
and this non-uniform lattice is thus well-defined.
Consider the following Hamiltonian:
\begin{equation}
H=-{1\over 2} \sum_{1\le i\ne j\le L} J_{ij} \Pi_{ij},
\label{eq:hamil}
\end{equation}
where $J_{ij}=1/(r_i-r_j)^2$, and the permutation
operator $\Pi_{ij}$ is given by
\begin{eqnarray}
\Pi_{ij}=&&c_{j\uparrow}^\dagger c_{i\uparrow} (1-n_{i\downarrow}
-n_{j\downarrow} ) + c_{i\uparrow}^\dagger c_{j\uparrow}
(1-n_{i\downarrow} - n_{j\downarrow} )\nonumber\\
&&+ c_{j\downarrow}^\dagger c_{i\downarrow} (1-n_{i\uparrow} -n_{j\uparrow})
+ c_{i\downarrow}^\dagger c_{j\downarrow}
(1-n_{i\uparrow} - n_{j\uparrow})\nonumber\\
&&+{1\over 2} (n_i-1)(n_j-1) + c_{i\uparrow}^\dagger
c_{i\downarrow}^\dagger c_{j\downarrow} c_{j\uparrow}
+c_{i\downarrow} c_{i\uparrow}
c_{j\uparrow}^\dagger c_{j\downarrow}^\dagger\nonumber\\
&&-{1\over 2} (n_{i\uparrow} - n_{i\downarrow})
(n_{j\uparrow} - n_{j\downarrow})\nonumber\\
&&-c_{i\downarrow}^\dagger c_{i\uparrow} c_{j\uparrow}^\dagger c_{j\downarrow}
-c_{i\uparrow}^\dagger
c_{i\downarrow} c_{j\downarrow}^\dagger c_{j\uparrow}\nonumber\\
&&+ (n_{i\uparrow} -1/2) (n_{i\downarrow} - 1/2)
+ (n_{j\uparrow} -1/2) (n_{j\downarrow} -1/2),
\label{eq:permu}
\end{eqnarray}
where $c_{i\sigma}$ and $c_{i\sigma}^\dagger$
are electron annihilation and creation
operators at site $i$ with spin $\sigma=\uparrow, \downarrow$.
The electron number operators are given by
$n_{i\uparrow} = c_{i\uparrow}^\dagger c_{i\uparrow},
n_{i\downarrow}=c_{i\downarrow}^\dagger c_{i\downarrow}, n_i=n_{i\uparrow}
+n_{i\downarrow}$.
If one defines the Hamiltonian on a uniform lattice and the coupling
parameter $J_{ij} =\delta_{1,|i-j|}$, the system becomes the extended
Hubbard model studied by Ebler, Korepin and Schoutens\cite{korepin2}.

For the long range extended Hubbard model defined on this non-uniform
lattice, we'll present its ground state, the energy spectrum and
the proof of integrability in the following. At each site $i$, one
can use the standard slave boson representation:
\begin{eqnarray}
&&|\downarrow\uparrow> <0| = p_i^\dagger b_i\nonumber\\
&&|\downarrow\uparrow> <\sigma| = p_i^\dagger f_{i\sigma}\nonumber\\
&&|\sigma> <\tau| = f_{i\sigma}^\dagger f_{i\tau}\nonumber\\
&&|\sigma> <0| = f_{i\sigma}^\dagger b_i \nonumber\\
&&|0><\sigma| = b_i^\dagger f_{i\sigma},
\end{eqnarray}
where $|\downarrow\uparrow>=c_{i\downarrow}^\dagger c_{i\uparrow}^\dagger |0>,
|\sigma>=c_{i\sigma}^\dagger |0>$, at site $i$.
For these relations to hold, the pair operators $p$ and the hole
operators $b$ are bosonic, while the operators $f$ are fermionic:
\begin{eqnarray}
&&[b_i,b_j^\dagger]=\delta_{ij}\nonumber\\
&&[p_i,p_j^\dagger]=\delta_{ij}\nonumber\\
&&[f_{i\sigma}, f_{j\tau}^\dagger]_+ = \delta_{ij} \delta_{\sigma\tau},
\end{eqnarray}
with the constrain that at any site $i$ there is always one particle, i.e.,
$\sum_{\sigma=\uparrow,\downarrow}
f_{i\sigma}^\dagger f_{i\sigma} + b_i^\dagger
b_i + p_i^\dagger p_i = 1$.
With the slave boson representation, a state vector can be written
as below:
\begin{eqnarray}
|\phi> =\sum_{\{x\}, \{y\}, \{z\}} \phi (x_1, x_2, \cdots, x_A |
&&y_1\sigma_1,y_2\sigma_2,\cdots,y_M\sigma_M
|z_1,z_2,\cdots,z_Q)\cdot \nonumber\\
&&\cdot \prod_{\alpha=1}^A p_{x_\alpha}^\dagger
\prod_{l=1}^M f_{y_l\sigma_l}^\dagger
\prod_{i=1}^Q b_{z_i}^\dagger |0>,
\label{eq:amplitude}
\end{eqnarray}
where the wavefunction $\phi$ is symmetric when exchanging $\{x\}$
or $\{z\}$ respectively, while anti-symmetric when exchanging
$y_i\sigma_i$ and $y_j\sigma_j$. In the following, we
use $(q_1, q_2, \cdots, q_L)=
(x_1, \cdots, x_A | y_1, \cdots, y_M |z_1, \cdots, z_Q)$.

With the Hamiltonian $H$ given by Eq.~(\ref{eq:hamil}), the number
of the electron pairs, the number of holes, the number of the sites
single-occupied by up-spin electrons, the number of the sites
single-occupied by the down-spin electrons, are all conserved quantities.
Let us denote them by the following notations:
\begin{eqnarray}
&&A=\sum_{i=1}^L n_{i\downarrow} n_{i\uparrow}\nonumber\\
&&M_\uparrow=\sum_{i=1}^L n_{i\uparrow}-A\nonumber\\
&&M_\downarrow=\sum_{i=1}^L n_{i\downarrow}-A\nonumber\\
&&Q=L-M_\uparrow - M_\downarrow - A\nonumber\\
&&M=M_\uparrow+M_\downarrow.
\end{eqnarray}
In the following, we work in the Hilbert space of fixed
$A, M_\uparrow, M_\downarrow$ and $Q$.

Using the amplitude $\phi$ defined
by the Eq.~(\ref{eq:amplitude}), the eigenenergy equation can thus
reduce to
\begin{equation}
-{1\over 2} \sum_{1 \le i\ne j \le L} {1\over (q_i-q_j)^2} M_{ij}
\phi(\{q\}) = E \phi(\{q\}),
\label{eq:first}
\end{equation}
where the operator $M_{ij}$ permutes the coordinates $q_i$ and $q_j$.
The Hamiltonian $H=-{1\over 2} \sum_{1\le i \ne j\le L} (q_i-q_j)^{-2} M_{ij}$
commutes with the infinite number of simultaneous constants of motion,
i.e., $[H, I_n]=0, [I_n, I_m]=0$, with $n,m=1, 2, \cdots, \infty$,
where $I_n=\sum_{i=1}^L h_i^n, h_i=a_i^\dagger a_i,$
with $a_j^\dagger=\sum_{k(\ne j)=1}^L i(q_j-q_k)^{-1} M_{jk} +i q_j,
a_i=(a^\dagger)^\dagger$, with $j=1,2,\cdots, L$\cite{poly,wang1}.
This shows that this long range version of extended Hubbard model is also
integrable.
When there are no electron pairs, this system reduces to the long range
t-J model studied before\cite{wang1,wang2}. When there are no electron pairs
and
no holes, the system becomes a previous spin chain\cite{poly}.

We first construct the ground state wavefunction in the general subspace
of fixed non-zero $A$, $M_\uparrow$, $M_\downarrow$ and $Q$. Let us look
at the following Jastrow product wavefunction:
\begin{eqnarray}
\phi (x_1, x_2, \cdots, x_A &&|y_1\sigma_1, y_2\sigma_2, \cdots, y_M\sigma_M
|z_1, z_2, \cdots, z_Q)\nonumber\\
&&= \prod_{1\le i<j\le M} (y_i-y_j)^{\delta_{\sigma_i\sigma_j}}
e^{i{\pi\over 2} sgn (\sigma_i - \sigma_j)}.
\end{eqnarray}
This wavefunction is anticipated to be the ground state of the system
in the subspace of fixed $A, M_\uparrow, M_\downarrow$ and $Q$.
With previous experience\cite{wang2}, one may show that this wavefunction is
indeed an eigenstate of the Hamiltonian, with the following eigenenergy
\begin{equation}
E_0=-{1\over 4} L(L-1) + {1\over 2} M_\uparrow (M_\uparrow -1) +
{1\over 2} M_\downarrow (M_\downarrow -1).
\end{equation}

For this Hamiltonian, one may show that $a_j^\dagger$ and $a_j$
are raising or lowering
operators defined on this lattice,
$[a_j^\dagger, H]=-a_j^\dagger, [a_j, H]=a_j$.
One may simply prove following identities:
\begin{eqnarray}
&&a_i \phi = 0; \,\,\,\, i=1, 2, \cdots, M\nonumber\\
&&a_i \phi = 0; \,\,\,\, i=A+M+1, A+M+2, \cdots, L.
\end{eqnarray}
Therefore, one can see that
\begin{eqnarray}
&&(\sum_{i=1}^M a_i) \phi = 0,\nonumber\\
&&(\sum_{i=A+M+1}^L a_i ) \phi =0.
\end{eqnarray}
Furthermore, one can show that
\begin{equation}
( \sum_{i=1}^M a_{i+A}\sigma_i^z ) \phi = 0.
\end{equation}
The above three identities show the impossibilities of constructing
non-zero eigenstates of lower energy than $E_0$ with the annihilation
operators. One may regard this as a partial support for our idea
that $\phi$ is the ground state in the subspace of fixed $A, M_\uparrow,
M_\downarrow, Q$. For a more complete confirmation, further work is needed
in this case.

The full energy spectra of the system consists of energy levels equal-spaced,
\begin{equation}
E(s)=E_0 + s,
\end{equation}
where $s=0, 1, 2, \cdots$. For a lattice of finite size $L$, there is an upper
bound on the value of $s$, as the Hilbert space of the system is finite.
There are several ways of constructing excited states. The first way is to
excite those electron pairs:
\begin{equation}
|n_1, n_2, \cdots, n_A> =\sum_P \prod_{i=1}^A (a_i^\dagger)^{n_{P_i}} |\phi>,
\end{equation}
where $n_1, n_2, \cdots, n_A$ are integers or zero, the summation $P$
is over all possible permutations. Under the operation of
the creation operators, any states constructed this way, if not vanishing,
will be the eigenstates of the Hamiltonian with eigenenergy
\begin{equation}
E_1=E_0 + (n_1+n_2+\cdots+n_A).
\end{equation}
The second way to create excitations is to excite the holes of the system:
\begin{equation}
|m_1, m_2, \cdots, m_Q> = \sum_P (a_{A+M+1}^\dagger)
^{m_{P_1}} (a_{A+M+2}^\dagger)^{m_{P_2}}
\cdots (a_{A+M+Q}^\dagger)^{m_{P_Q}} |\phi>,
\end{equation}
with the eigenenergies given by
\begin{equation}
E_2=E_0+ (m_1+m_2+\cdots+m_Q),
\end{equation}
if these state vectors are not zero.
Finally, one may create excitations by exciting the electrons on the
sites single-occupied. These excited states may be written as
\begin{equation}
|\phi_3> = \sum_P (S_1(\nu_1)^\dagger)^{n_{P_1}}
(S_2(\nu_2))^\dagger)^{n_{P_2}} \cdots (S_M(\nu_M)^\dagger)^{n_{P_M}} |\phi>,
\end{equation}
where $S_i(\nu_i)^\dagger=a_{i+A}^\dagger \sigma_i^{\nu_i}$, with
$\nu_i=0, \pm, z$. These states have eigenenergies given by
\begin{equation}
E_3=E_0+ (n_1+n_2+\cdots+ n_M ).
\end{equation}
Clearly, excited energy levels are highly degenerate.
However, we are still unable to develop a systematic rule to
characterize the pattern of the degeneracy and to explain
it with symmetries of the system (presumably, Yangian
symmetry).

In the limiting case where $M=M_\uparrow+M_\downarrow=0$, one has
two types of bosons on the chain, the local electron pairs
and the holes. The wavefunction $\phi$ becomes
\begin{equation}
\phi(x_1,x_2,\cdots,x_A|z_1,z_2,\cdots,z_Q)=1,
\end{equation}
which is obviously the ground state of the Hamiltonian $H$, with
the ground state energy $E_0=-L(L-1)/4$.
With the raising operators, one may construct the first excited
state quite easily:
\begin{equation}
\phi'(x_1,x_2,\cdots,x_A|z_1,z_2,\cdots,z_Q)=(x_1+x_2+\cdots+x_A).
\end{equation}
The wavefunctions of higher excited states may be obtained by
acting this wavefunction with raising operators again.

Besides this extended Hubbard model of long range type defined
on a non-uniform lattice, we would like to introduce another
model of extended Hubbard model of long range type.
Consider a uniform lattice of sites equal-spaced. The size of the lattice
is $L$. The Hamiltonian of the system is given by
\begin{equation}
H=-{1\over 2} \sum_{1\le i \ne j \le L} d(i-j)^{-2} \Pi_{ij},
\end{equation}
where the function $d(i-j)={L\over \pi} \sin(\pi(i-j)/L)$, and the permutation
operator $\Pi_{ij}$ is the same as Eq.~(\ref{eq:permu}).
With our previous experience, one can simply write the first
quantized Hamiltonian as follows:
\begin{equation}
H=-{1\over 2} \sum_{1\le i\ne j\le L} 1/d^2(q_i-q_j) M_{ij}.
\end{equation}
It is very easy to prove that this extended Hubbard model is
also integrable, once the Hamiltonian is written this way by using
the relation of Fowler and Minahan\cite{fowler}:
\begin{eqnarray}
&&L_n=\sum_{i=1}^L \pi_i^n, \, \pi_i=
\sum_{j(\ne i)=1}^L z_j/(z_i-z_j) M_{ij},\nonumber\\
&& [L_n, L_m]=0,~ [L_n, H]=0; ~\, n, m=1, 2, \cdots,
\end{eqnarray}
where $z_j=e^{2\pi i q_j/L}$, $j=1, 2, \cdots, L$.
This thus provides a proof for the integrability of the
extended Hubbard model. In the limiting case of
no electron pairs, this system
reduces to the long range t-J model\cite{kura,kawa,wang3,wang1}.
If there are no holes and no electron pairs, the Hamiltonian
reduces to the Haldane-Shastry spin chain\cite{haldane,shastry}.
In the general case,
to write the conserved quantities of the system in terms
of second quantization, one may
simply use the permutation symmetries of the wavefunction.
In this paper, we shall not
present the wavefunctions and the eigenenergy
spectrum of this system. More results on this system will be
given elsewhere.

In summary, we have introduced two integrable models of strongly
correlated electrons, one defined on a non-uniform lattice, the
other on a uniform lattice. The systems belong to
the Jastrow-integrable-type.
Particularly, the system on
the non-uniform lattice has equal-spacing energy spectrum.
Because of the degrees of freedom of the paired electrons,
the energy levels will have much larger degeneracy than
the long range t-J model. One of the most interesting
things is to determine the underlying symmetries explicitly that
account for the degeneracy of each energy level.

I wish to thank V. E. Korepin for interesting conversations.
Discussions with Mo-lin Ge and C. Gruber are acknowledged gratefully.
This work was supported by the Swiss National Science Foundation.

\end{document}